# Engineering the point spread function of layered metamaterials

A. PASTUSZCZAK, M. STOLAREK, and R. KOTYŃSKI*

University of Warsaw, Faculty of Physics
Pasteura 7, 02-093 Warsaw, Poland

*Layered metal-dielectric metamaterials have filtering properties both in the frequency domain and in the spatial frequency domain. Engineering their spatial filtering response is a way of designing structures with specific diffraction properties for such applications as sub-diffraction imaging, supercollimation, or optical signal processing at the nanoscale. In this paper we review the recent progress in this field.*

*We also present a numerical optimization framework for layered metamaterials, based on the use of evolutionary algorithms. A measure of similarity obtained using Hölder's inequality is adapted to construct the overall criterion function. We analyze the influence of surface roughness on the quality of imaging.*

## 1. Introduction

A decade after the introduction of the superlens [1,2] and its experimental demonstration [3,4], planar metal-dielectric metamaterials (MDM) have become a well understood element for imaging in the near-field. Superlenses and other layered metamaterials amplify evanescent waves. Therefore, in principle, they are not limited by the classical diffraction limit that results from low-pass spatial filtering of the spatial spectrum occurring during propagation of plane waves in free space [5]. More generally, all stratified optically linear media (including uniform dielectrics, Fabry-Perot (FP) etalons, periodic or aperiodic dielectric or metal-dielectric multilayers, and homogenized effective medium stacks) are frequency filters and spatial filters at the same time. In this paper we are interested in the spatial filtering only, and we assume a monochromatic illumination. In the optical wavelength range light is normally not transmitted through metallic slabs thicker than the skin-depth, which is of the order of 10-20 nm for noble metals. This thickness limitation can be mitigated using the mechanism of resonant tunneling, which enables light to be transmitted through a double barrier, with theoretical transmission of 100% in the lossless case [6]. Such a value cannot be retained in the presence of absorption or scattering; nevertheless, metal-dielectric stacks can be largely transparent even when they contain a substantial proportion of metal [7,8]. A metal-dielectric multilayer consisting of very thin layers is equivalent to a material characterized with the effective dispersion relation of a uniaxial crystal [9]. In practice, the effective medium approximation provides a qualitatively accurate permittivity model for optical wavelengths when the layer thickness is of the order of 10 nm. It is also known that the effective medium approximation tends to overvalue the losses [10,11]. The effective medium is equivalent to an anisotropic material, which may have an extreme, theoretically infinite, extraordinary permittivity. The huge birefringence can be used for diffraction-free, super-collimating, and sub-diffraction guidance of light. Additionally, as a whole, the multilayer forms an etalon with effective permittivity, providing maximal transmission when the total thickness creates FP resonances [12,13]. When the ordinary permittivity equals that of the surrounding medium, thanks

---

to impedance matching, the reflections from the structure are small and the role of FP resonances is less important. Such a metamaterial consisting of the effective medium can be arbitrarily shaped [14]. Negative refraction, sub-wavelength focusing, tailored diffraction, and sub-wavelength imaging are some of the effects attainable with layered metamaterials [15-19]. The dispersion relation of the effective medium with an elliptically shaped iso-frequency curve is suitable for obtaining sub-wavelength imaging with good fidelity. On the other hand, materials with an opposite sign of the effective permittivity tensor components show peculiar properties resulting from the hyperbolic dispersion relation [9,20] and have been utilized to construct hyperlenses [21] capable of magnifying sub-diffraction objects to images which can be further viewed with microscopic techniques. More recently, a high performance absorber based on a hyperbolic material has been theoretically investigated [11]. The third kind of dispersion relation is exhibited by the zero-permittivity metamaterials [22]. These materials, which may be seen as an intermediate case between hyperbolic and elliptical materials, are important for optical cloaking [23] or funneling [24]. In practice, due to the imaginary part of the effective permittivity, real layered materials can only approximately be attributed to one of these three groups.

## 2. Transfer function of layered metamaterials

Transmission of monochromatic light through a layered metamaterial slab is a linear spatial filtering operation. The same can be said about transmission through any other stratified optically linear structure, and also about propagation through free space. Therefore, the linear system theory is suitable for describing the imaging properties of stratified optically linear structures. The transfer function of a layered metamaterial can be used to determine the resolution of the metamaterial, as well as imaging artifacts, aberrations, and the transmission coefficient. Assuming that the boundaries between different layers are situated along the planes $z = const$, the one-dimensional spatial spectrum of the magnetic field $H_y$, in the planes $z = const$ is equal to

$$\widehat{H}_y(k_x, z) = \int_{-\infty}^{\infty} H_y(x, z) \cdot \exp(ik_x x) \cdot dx. \tag{1}$$

We have chosen $H_y$ here, since this field component is continuous within the entire planar system for the TM polarization. For the TE polarization, $H_y$ should be replaced with $E_y$.

In a uniform material, the transfer function between $z = 0$ and $z = L$ is equal to

$$\hat{t}(k_x) = \frac{\widehat{H}_y(k_x, z=L)}{\widehat{H}_y^{inc}(k_x, z=0)} = \exp(ik_z L) = \exp\left(iL\sqrt{k_0^2 n^2 - k_x^2}\right). \tag{2}$$

This transfer function has a cutoff at $k_x = k_0 n$ and, as is well known from Fourier Optics, it defines a low-pass spatial filter and puts a limit on the spatial resolution of the system. In a layered or stratified medium, the transfer function can be calculated using the transfer matrix method.

Extension to two-dimensional filtering is possible [25] but is less straightforward than in scalar wave optics.

In [26] it has been shown that, for anisotropic media with indefinite sign of the permittivity or permeability, there might exist a cutoff of the transfer function; however, an anti-cutoff may exist instead. The dispersion relation of a uniaxial material is defined with two different formulas for the TE and TM polarizations:

$$k_z^{TE} = \pm\sqrt{\epsilon_y\mu_x k_0^2 - \frac{\mu_x}{\mu_z}k_x^2}, \tag{3a}$$

$$k_z^{TM} = \pm\sqrt{\epsilon_x\mu_y k_0^2 - \frac{\epsilon_x}{\epsilon_z}k_x^2}. \tag{3b}$$

The transfer function still has the form $\hat{t}(k_x) = \exp(ik_z L)$, and the conditions for the existence of a cutoff or anti-cutoff are

$$\text{TE cutoff: } \epsilon_y\mu_x > 0, \frac{\mu_x}{\mu_z} > 0, \text{ TE anticutoff } \epsilon_y\mu_x < 0, \frac{\mu_x}{\mu_z} < 0 \tag{4a}$$

$$\text{TM cutoff: } \epsilon_x\mu_y > 0, \frac{\epsilon_x}{\epsilon_z} > 0, \text{ TM anticutoff } \epsilon_x\mu_y < 0, \frac{\epsilon_x}{\epsilon_z} < 0 \tag{4b}$$

If the metamaterial consists of thin metallic and dielectric layers without magnetic properties, according to the effective medium theory, the effective permittivity tensor has principal components given by the arithmetic and harmonic means of the permittivities $\epsilon_i$ of the layers

$$\epsilon_x = \epsilon_y = (\Sigma_i d_i \epsilon_i)/(\Sigma_i d_i), \tag{5a}$$

$$\epsilon_z = (\Sigma_i d_i)/(\Sigma_i d_i/\epsilon_i), \tag{5b}$$

where $i$ enumerates the nearest layers over which the structure is assumed to be homogenized (e.g. the layers within one period of a periodic stack), and $d_i$ are the corresponding layer thicknesses.

Let us assume that the metamaterial is periodic and consists of two materials (i.e. $i = 1,2$), one of which is a metal, in our case silver, and the other is a dielectric. The layered metamaterial is shown schematically in Fig. 1.

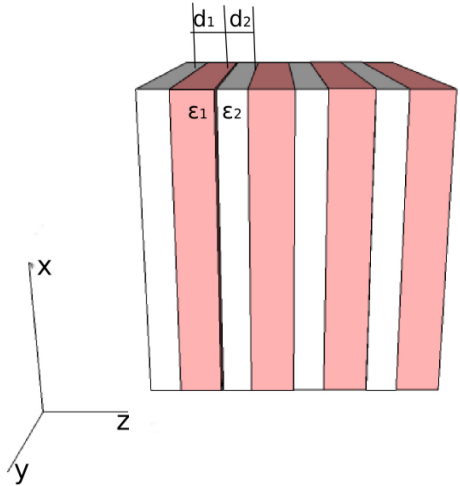

Fig. 1: Schematic of the metal-dielectric metamaterial

For the purpose of simulations, the permittivity of silver is commonly taken from [28] (which we will use further) or from [29]. We would like to point out that the dispersion and losses of silver in the visible range taken from these two sources show significant differences that originate from the way in which the samples were prepared. In Fig. 2a we present the dispersion of the real and imaginary part of the permittivity of silver from [28] and from [29], respectively. Subsequently, in Fig. 2b we show the corresponding effective skin depth (or penetration depth), defined as $\delta =$

$\lambda/2\pi \text{Im}(\sqrt{\epsilon_x})$ for a multilayer, which is almost free from diffraction thanks to the extreme effective birefringence. This multilayer consists of silver layers with permittivity $\epsilon_1$ and layer thickness $d_1 \ll \lambda$, and of dielectric layers with permittivity $\epsilon_2 = 1 - \text{Re}(\epsilon_1)$ and thickness $d_2 \ll \lambda$ such that $d_1/d_2 = \text{Re}(\epsilon_1)/\epsilon_2$. As we see from Fig. 2b, the effective skin depth calculated using dispersion data from [28] and [29] differs by a factor of 2-4. The scale of this difference has a profound effect on the design strategy of MDM in any particular application.

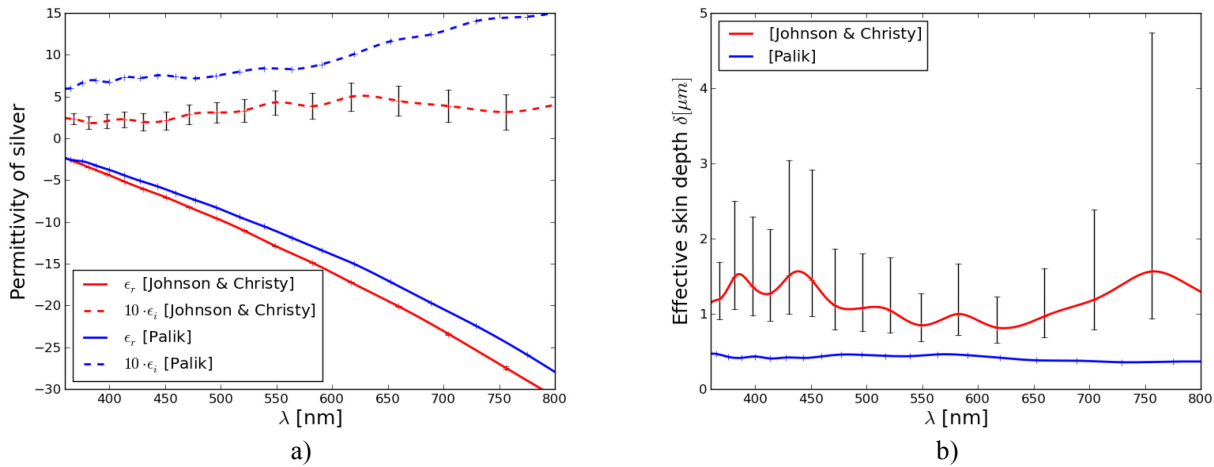

Fig. 2: a) Permittivity of silver after Johnson and Christy, Ref. [28], and Palik, Ref [29]; b) theoretical effective skin depth of a diffraction-free silver-dielectric MDM superlens.

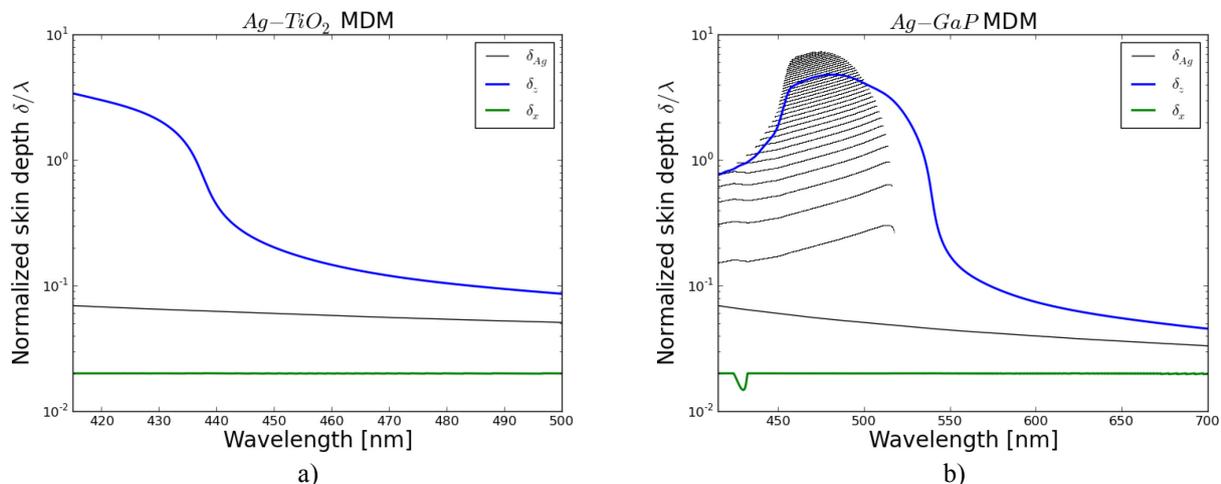

Fig. 3: Skin depth of silver, effective skin depth of an infinite MDM, and SPP range as a function of wavelength for a) Ag-TiO$_2$ MDM; b) Ag-GaP MDM. Black points indicate the MDM widths which satisfy an FP resonance for transmission.

In the rest of this section we focus on realistic MDMs that consist of silver and either TiO$_2$ (a low refractive index dielectric) or GaP (a high refractive index semiconductor). The permittivities of TiO$_2$ and GaP are taken from [29]. At first, we assume the validity of the effective medium model. This is justified when the layers are thin, i.e. when $d_i \ll \lambda$. Some insight into the imaging properties of the MDM can be deduced from the penetration depths $\delta_x = \lambda/2\pi \text{Im}(\sqrt{\epsilon_z})$, $\delta_z = \lambda/2\pi \text{Im}(\sqrt{\epsilon_x})$. The depth $\delta_z$ is a rough estimation of the limit of the thickness of the MDM introduced by losses, while $\delta_x$ provides an estimation of the resolution of MDM. In Fig. 3 we show the dispersion of $\delta_x$ and $\delta_z$, comparing them to the skin depth of bulk silver $\delta_{Ag}$ for an MDM that minimizes the heuristic criterion $E_0 = max(\delta_x, \lambda/50)/min(\delta_z, 10\lambda)$. As we can see, the skin depth of the MDM may be larger than $\delta_{Ag}$ by as much as two orders of magnitude, while the resolution remains deeply sub-wavelength $\delta_x \sim \lambda/50$.

The transfer function of the MDM depends not only on its effective dispersion, but also on the FP structure of the MDM seen as a slab surrounded by an external medium, such as air. Thicknesses corresponding to FP resonances on the MDM slab are indicated in Fig. 3b. At these thicknesses of the entire MDM, the transmission is maximal. In the regions of Fig. 3ab without FP resonances, transmission decreases monotonically with the thickness of MDM.

The transfer function of an MDM which supports superresolution should be approximately constant for the spatial frequencies up to some value $k_x/k_0>1$. In a strongly birefringent material, the FP resonances have a weak dependence on $k_x/k_0$ and one can obtain resonant transmission for a wide range of spatial frequencies. The transfer function of the Ag-GaP MDM, calculated within the effective medium approximation, is shown in Fig. 4, as a function of the thickness of MDM. The sub-plots correspond to the three wavelengths 450 nm, 480nm, and 550 nm, with the second one allowing for the largest number of FP resonances to be supported (See Fig. 3b). Notably, the flat shape of the transfer function for this wavelength is an indication that the corresponding point spread functions (PSF) may be expected to have a sub-wavelength size. The PSF for two selected resonances are presented in Fig. 5, where we also show the effect of a finite size of the pitch. As compared to the effective medium model, taking into account the finite size of the pitch deteriorates the resolution and transmission efficiency, but the PSF remains sub-wavelength in size.

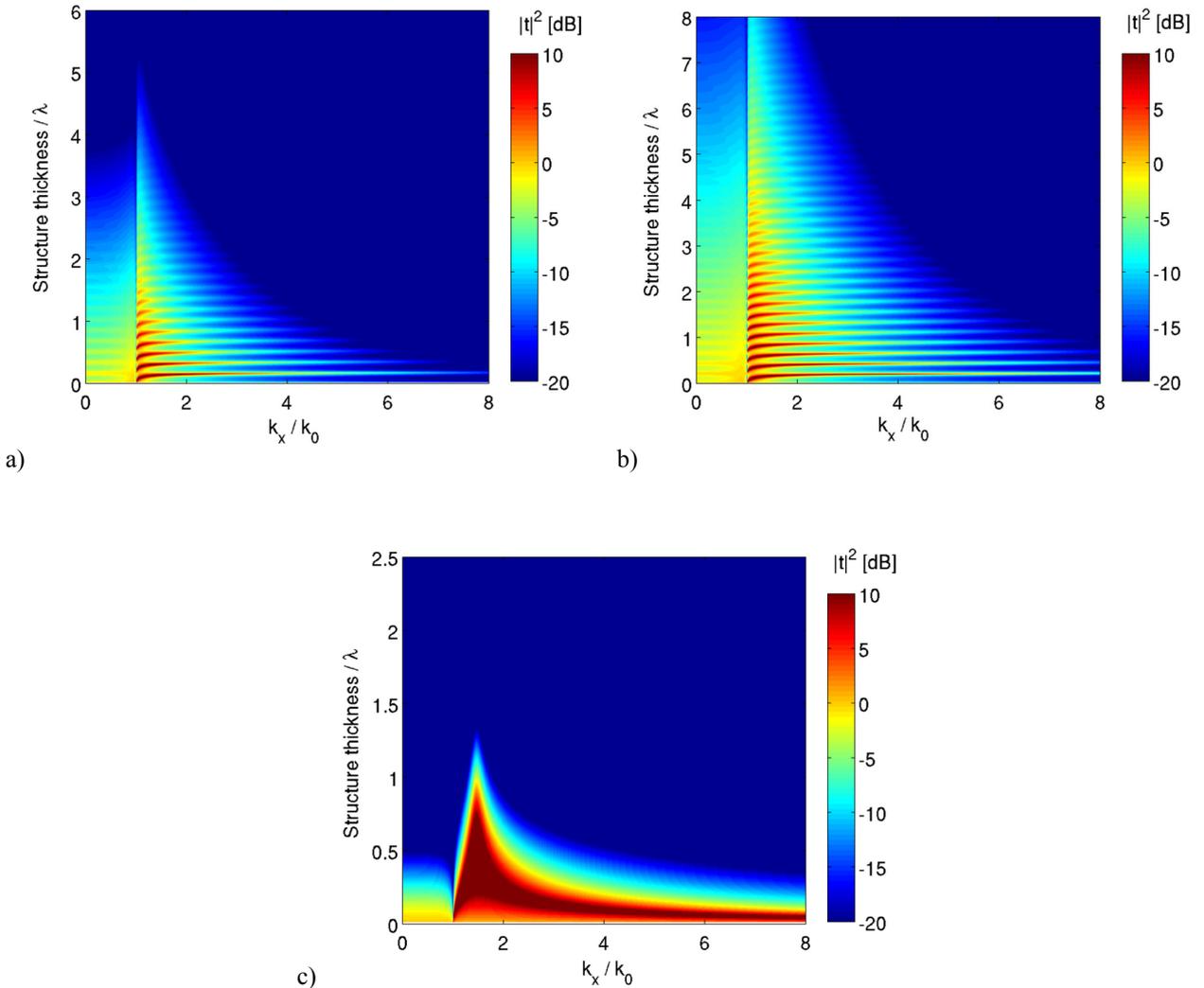

Fig. 4: Transfer function of the Ag-GaP MDM in the effective medium approximation as a function of the thickness of the MDM. The wavelength is equal to a) 450 nm, b) 480 nm, c) 550 nm.

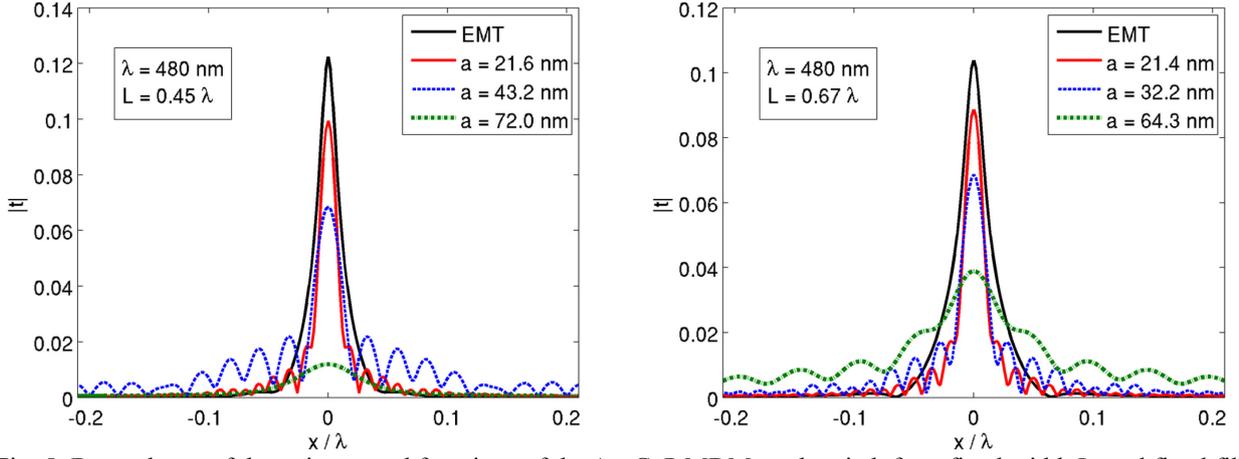

Fig. 5: Dependence of the point spread functions of the Ag-GaP MDM on the pitch for a fixed width L, and fixed filling fraction.

As we have mentioned, the simulations presented in Figs. 3-5 have been performed with optimized MDMs. The filling ratio $d_{Ag}/\Lambda$, where $d_{Ag} = d_1$ and $\Lambda = d_1 + d_2$ is the pitch, resulting from MDM optimization is presented in Fig. 6. The dispersion of the corresponding effective refractive index $n_o^{EMT}$ and extinction index $\kappa_o^{EMT}$, defined as $n_o^{EMT} + i\kappa_o^{EMT} = \sqrt{\epsilon_x}$, are presented in the same figure. Finally, while $\kappa_o^{EMT}$ is closely related to $\delta_z$, the effective index $n_o^{EMT}$ may be tuned to obtain impedance matching with the surrounding medium and at the same time to reduce the strength of FP resonances.

The effective medium, apart from anisotropy, may also have a different sign of the principal elements of the permittivity tensor. Within a certain wavelength range, denoted with dark blue color in Fig. 6, the signs of real parts of $\epsilon_x$ and $\epsilon_z$ become opposite and the dispersion relation (3b) is hyperbolic.

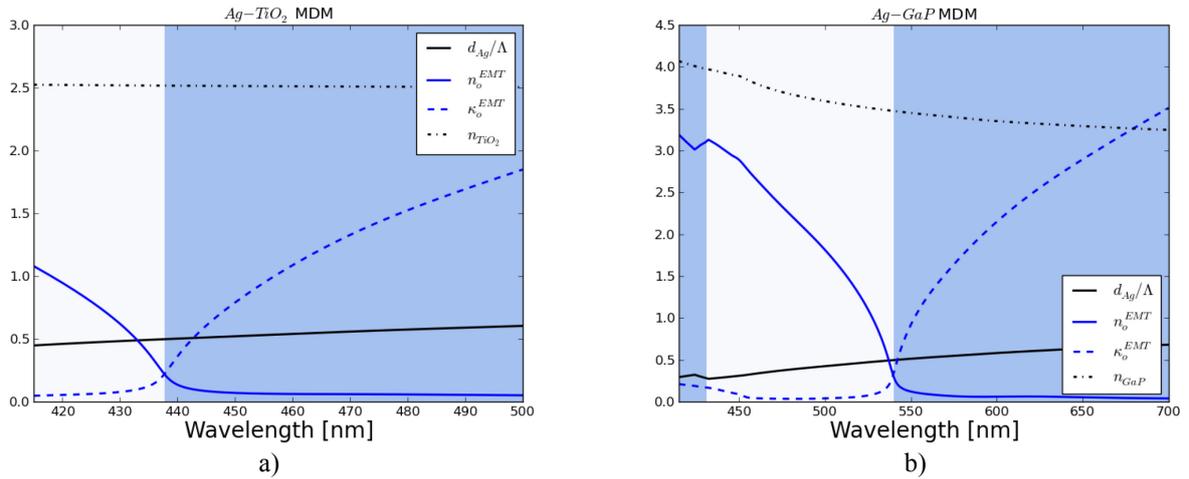

Fig. 6: Wavelength dependence of the filling fraction, of the effective refractive index, and of the effective extinction coefficient, of a diffraction-free MDM. The MDM consists of Ag and $TiO_2$ (a) or of Ag and GaP (b). The spectral region with hyperbolic dispersion is shown in blue (on-line).

The perpendicular component of the effective permittivity $\epsilon_z$ experiences a resonance at wavelengths where the dispersion relation changes from hyperbolic to elliptical. This resonance is illustrated in Fig. 7, as a function of the filling factor. In the vicinity of the resonance, the MDM becomes strongly birefringent and can be used for supercollimation.

Homogenization of the MDM based on the effective medium theory is only valid when the layers are thin, in practice of the order of $\lambda/10$-$\lambda/50$. For an infinite and periodic MDM the effective index and extinction index may be calculated from the Bloch wavevector of the periodic stack $k^{Bloch}$, as $n_o^{Bloch} + i\kappa_o^{Bloch} = k^{Bloch}/k_0$ [30]. Notably, $\kappa_o^{Bloch}$ does not depend on the choice of the Brillouin zone, and $n_o^{Bloch} + i\kappa_o^{Bloch}$ converges to $n_o^{EMT} + i\kappa_o^{EMT}$ when $\Lambda \to 0$ and the filling ratio is kept constant. We will now show that the extinction coefficient $\kappa_o^{EMT}$ predicted by the effective medium theory may be significantly larger than $\kappa_o^{Bloch}$, even when the pitch is subwavelength $\Lambda \sim \lambda/10$. This means that the effective medium theory tends to overvalue the losses. For the purpose of example, in Fig. 8 we present the dependence of $\kappa_o^{Bloch}$ on the filling fraction $d_{Ag}/\Lambda$ for MDMs with a finite size of pitch. Figures 8a, b, and c are calculated for the same wavelengths as Figs. 7a, b, and c. With the pitch increased from zero upward, first the extinction index decreases, and subsequently cavity modes are created in individual dielectric layers. There exists some optimal value of the pitch at which the losses are minimal.

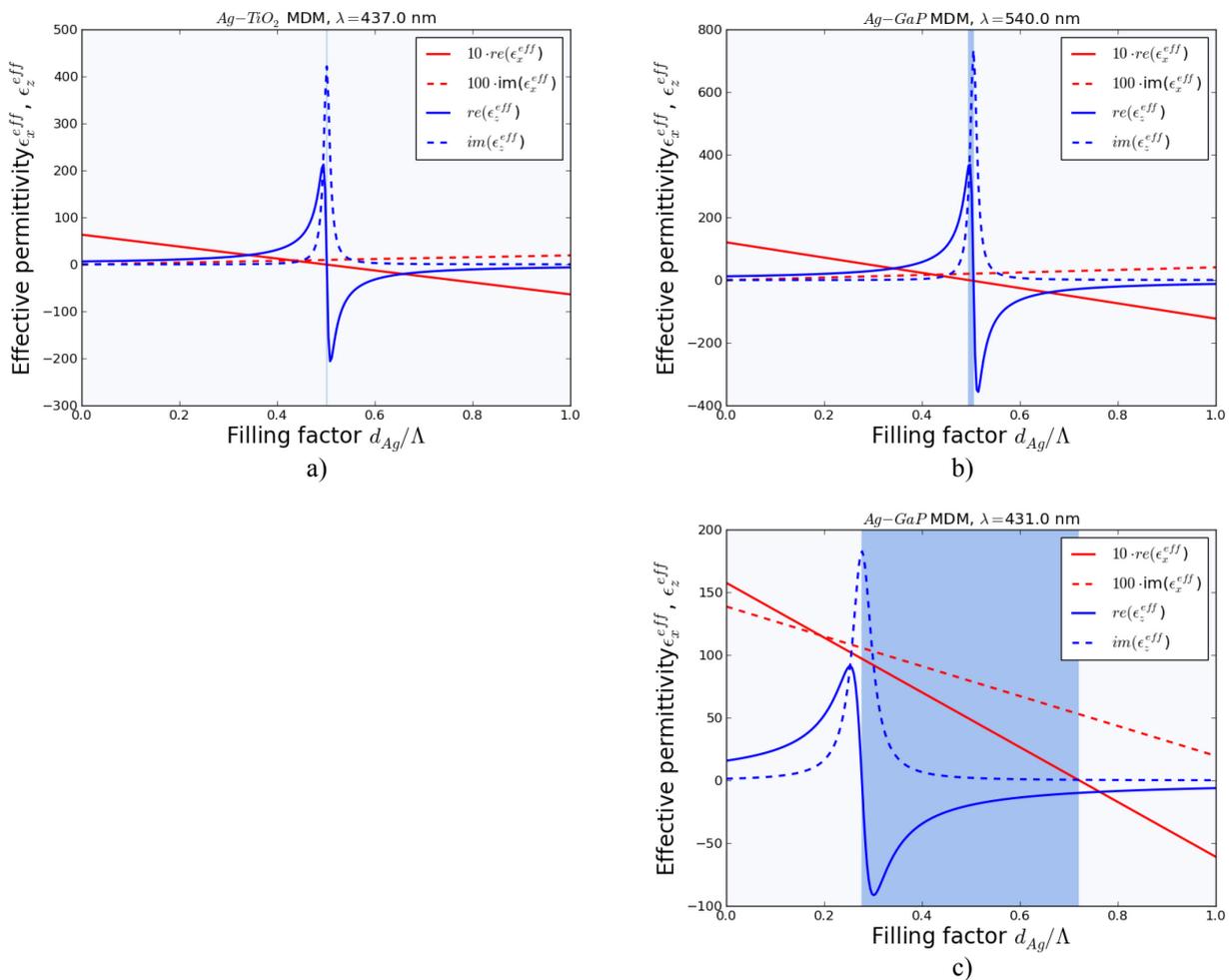

Fig. 7: Effective permittivity of an MDM as a function of filling fraction of silver. a) Silver-TiO$_2$ MDM at a wavelength of 437 nm; b) silver-GaP MDM at 431 nm; c) silver-GaP MDM at 540 nm. The region with hyperbolic dispersion is shown in a dark color (dark blue on-line).

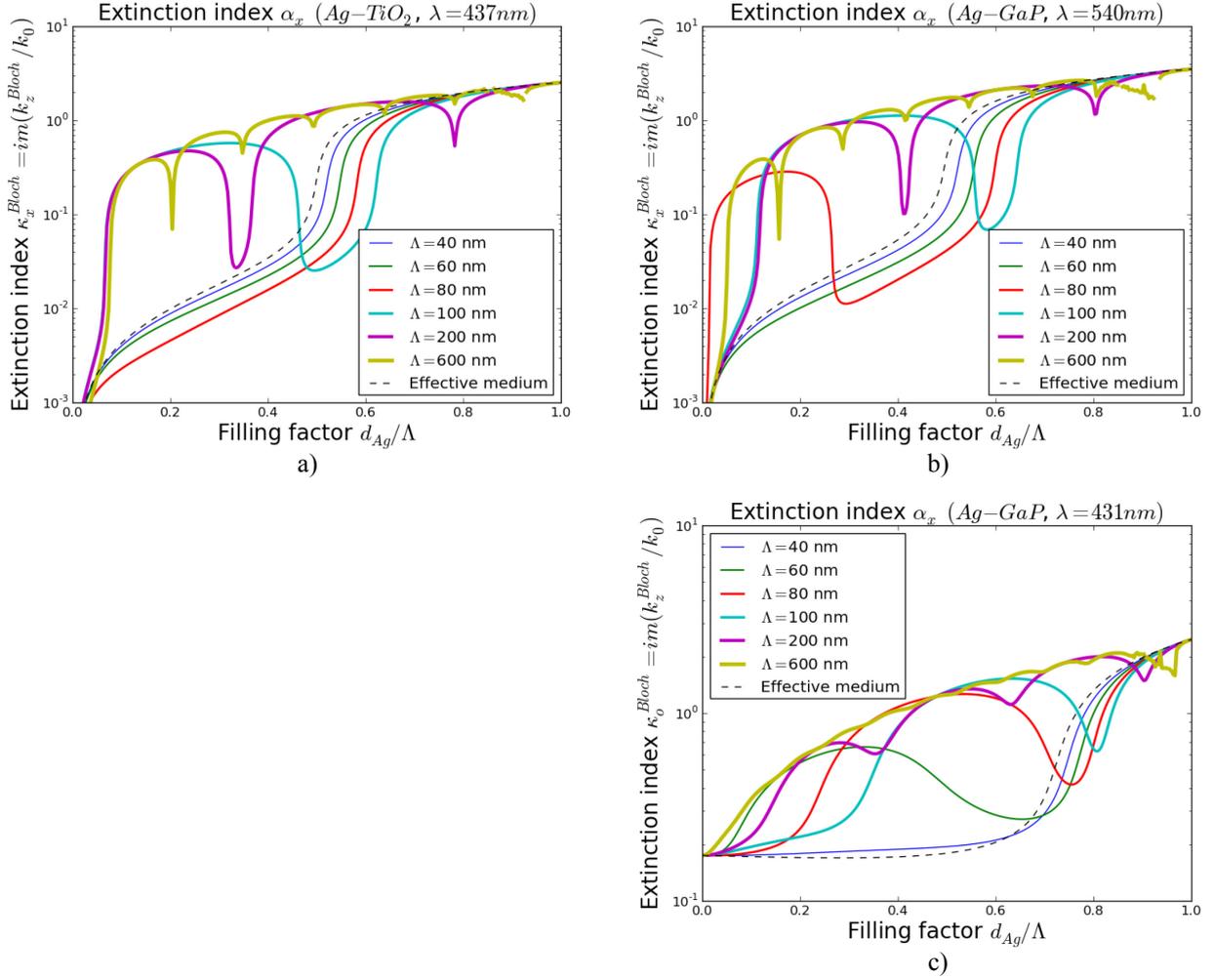

Fig. 8: Extinction ratio of the Bloch refractive index of an MDM with a varied period Λ, as a function of filling factor. a) Silver-TiO$_2$ MDM at 437 nm; b) Silver-GaP MDM at 540 nm; c) Silver-GaP MDM at 431 nm.

## 3. Optimization of MDM transfer function

Let us begin by introducing a similarity measure $G$ between two functions $t(k)$ and $r(k)$, which would be invariant to multiplication by any non-zero complex factor. The discretized functions will be written as vectors $t$ and $r$. These two functions refer to the transfer function of the MDM, and to the desired ideal transfer function that we intend to obtain, respectively. We adapt the expression for $G$ from intensity-invariant pattern recognition applications [27]. The measure is defined as

$$G(t,r) = \frac{\left|\sum_k w_k \cdot |t_k|^{\alpha-1} \cdot |r|^{\beta-1} r_k^* \cdot t_k\right|}{\left|\sum_k w_k \cdot |t_k|^{\alpha+\beta}\right|^{\alpha/(\alpha+\beta)} \cdot \left|\sum_k w_k \cdot |r_k|^{\alpha+\beta}\right|^{\beta/(\alpha+\beta)}}, \quad (6)$$

where $\alpha, \beta > 0$ and $w_k > 0$. It can be easily found that $G(t,r) \geq 0$. From Hölder's inequality, it can be shown that $G(t,r) \leq 1$. Moreover, $G(t,r) = 1$ if and only if the vectors $t$ and $r$ are collinear. The vector $w$ includes the weights that might vary for different spatial frequency $k = k_x/k_0$, and the coefficients $\alpha$ and $\beta$ weight the relative importance of amplitudes and phase in the compared functions.

In the following part of the paper, we will assume that the MDM consists of silver and TiO$_2$, and that the operating wavelength $\lambda$ is in the range between 400 nm and 500 nm. In case of imaging with sub-wavelength resolution, we additionally demand that the thickness of the filter is equal to at least 500 nm. The MDM consists of external TiO$_2$ layers of thickness $d_0$, a periodic stack of $N+1$ silver layers of thickness $d_{Ag}$, and of $N$ TiO$_2$ layers having thickness $d_x$. These assumptions are somehow arbitrary and are introduced to avoid obtaining trivial solutions or solutions which could not be compared with each other. Therefore, the optimized parameters are $(d_0, d_x, d_{Ag}, N, \lambda)$ with an additional constraint on the total thickness.

## 3.1 Imaging with sub-wavelength resolution

Sub-wavelength imaging with MDM metamaterials has been thoroughly investigated in recent years [1-4,8-10,12-16,21,22,25,30]. We propose the following set of criteria for the optimization of an MDM: i) the transmission $T(k_x)$ should be as large as possible for a broad range of angles of incidence; therefore, one of the criteria is $\langle T \rangle$, where the averaging takes place for $k_x/k_0$ in the range of [0,1]; ii) the second criterion is $\langle 1 - R \rangle$, where $R$ is the intensity reflection coefficient and the averaging takes place for $k_x/k_0$ in some selected subrange of the range [0,1]; and iii) the desired transfer function is equal to a constant $r = const(k_x)$. A constant transfer function corresponds to a delta-shaped point spread function (PSF). Notably, it does not matter what the amplitude and phase of this function is, as long as it is constant and the transmission is large enough. At the same time, it is advantageous to define the criterion with respect to the transfer function rather than PSF because an additional FFT operation is needed to calculate the PSF. We have assumed that the polarization is TM. Now, let us define an overall criterion that depends on $G$, $\langle T \rangle$, and $\langle R \rangle$ as $E(t,r) = \theta_1 \cdot G + \theta_2 \cdot \langle T \rangle + \theta_3 \cdot \langle 1 - R \rangle$, where $\theta_1 + \theta_2 + \theta_3 = 1$. Since $0 \leq G \leq 1$, $0 \leq \langle T \rangle \leq 1$, and $0 \leq \langle R \rangle \leq 1$, we have $0 \leq E \leq 1$. Moreover, the optimization results may be parameterized with $\theta_1, \theta_2, \theta_3$ resulting in a surface of optimal trade-offs of the three criteria considered.

We have optimized the MDM assigning equal weights to the three criteria $\theta_1 = \theta_2 = \theta_3 = 1/3$, and taking $w_{kx} = 1$ for $k_x/k_0 < 2$, and either $\alpha = \beta = 2$ or $\alpha = \beta = 0.1$. The optimized transfer functions are presented in Figs. 9a and b. Notably, a very flat phase characterizes both transfer functions. The width of the PSF (full width at half maximum of the squared modulus of the PSF for field component $H_y$) is equal to FWHM=$0.13\lambda$ and FWHM=$0.16\lambda$, which is a challenging result, especially since it is combined with a high transmission. The measure of similarity, $G$, is also very high in both cases and is approximately equal to 0.99, although its value cannot be directly compared when the parameters $\alpha, \beta$ are varied. The Finite-Difference-Time-Domain (FDTD [31]) simulation presented in Figs. 9c and d provides further insight into the diffraction-free sub-wavelength nature of imaging through the optimized MDMs. These figures show the $S_z$ component of the Poynting vector, while Figs. 9e and f show the cross-section of $S_z$ along the borders of MDM. The FWHM of the intensity profiles at these borders differ by approximately $\lambda/10$, which is consistent with the widths of PSF, and the diffraction-free nature of propagation through MDM.

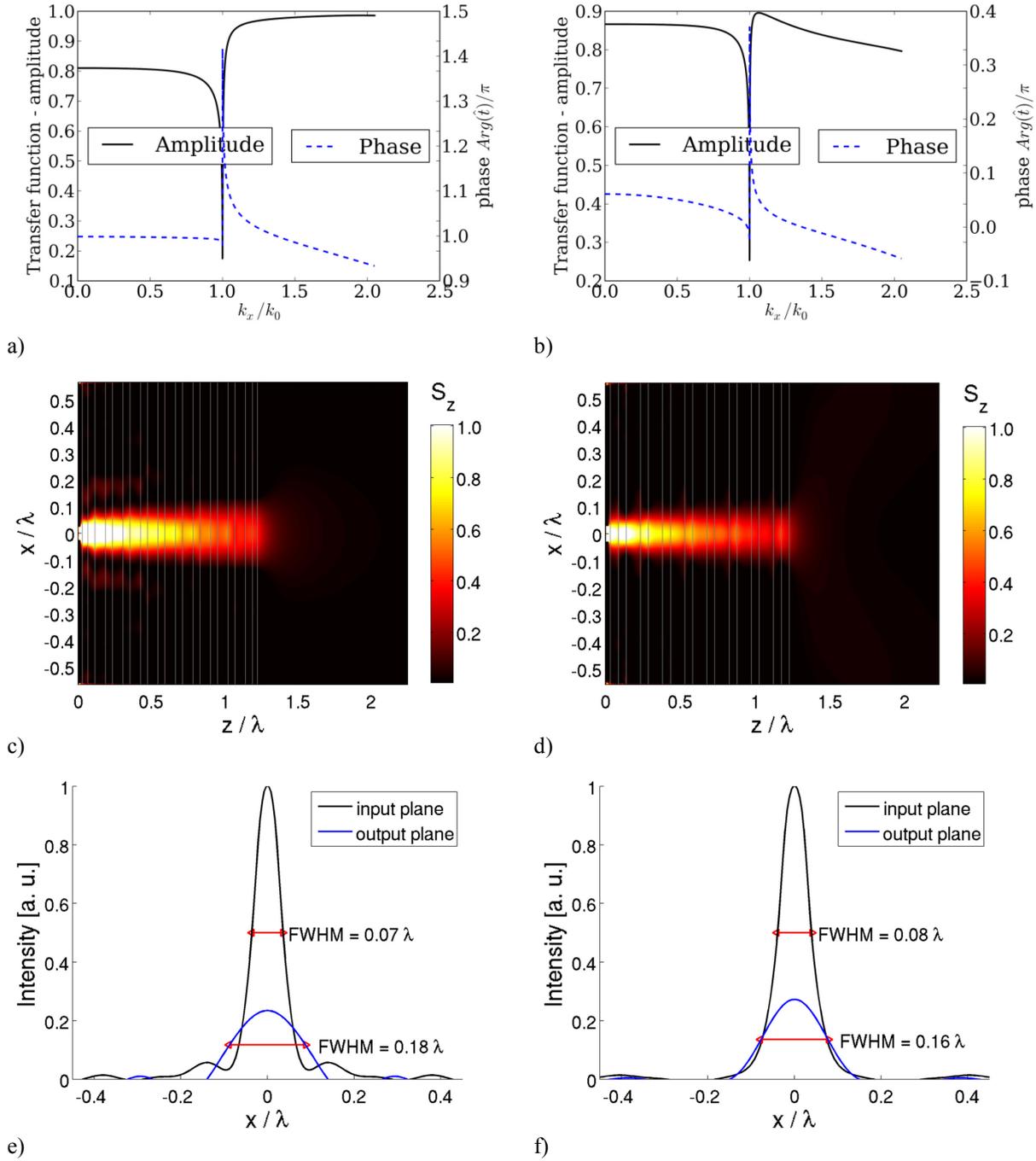

Figure 9: MDM optimized for sub-wavelength imaging. Transfer function (a,b), Poynting vector component $S_z$ (c,d), and intensity profiles (e,f) at the input and output sides of MDM (a,c,e). MDM is defined by the following parameters $d_0 = 14{,}5$ nm, $d_x = 29$ nm, $d_{Ag} = 21$ nm, $N = 9, \lambda = 427$ nm. The performance criteria are equal to $G = 0.987, \langle T \rangle = 0.625, \langle 1 - R \rangle = 0.990$, and $\alpha = \beta = 2$. b,d,f) MDM defined as $d_0 = 22$ nm, $d_x = 40$ nm, $d_{Ag} = 22$ nm, $N = 7, \lambda = 418$ nm. The performance criteria are equal to $G = 0.992, \langle T \rangle = 0.726, \langle 1 - R \rangle = 0.983$, and $\alpha = \beta = 0.1$.

## 3.2 Fresnel diffraction compensation

In this section we consider filters that compensate diffraction experienced by a wavefront propagating at a small distance. A plane wave propagating in air or in some dielectric material is subject to a change of phase that depends on the angle of propagation, the refractive index $n$, and the distance $d$. Therefore, the transfer function has a unit-magnitude, and a phase dependent on $k_z$, and it is equal to $t(k_x) = \exp(ik_z d) = \exp(id\sqrt{n^2 k_0^2 - k_x^2})$. At this point we neglect the evanescent part of the spatial spectrum. A Taylor expansion of $t(k_x)$ results in the expression for Fresnel diffraction. Now, let us consider a situation when the distance $d$ is small, of the order of one wavelength. It is possible to compensate for diffraction in this case using an MDM filter, for the TM polarization. In fact, the original Pendry's perfect lens consisting of a *40* nm silver slab is an example of such a filter. However, we are interested in compensating propagation effects at a distance larger than just $\lambda/10$ rather than on enhancing evanescent waves. For this purpose we have optimized an MDM layered filter, assuming that we want to compensate propagation at a distance of *200* nm. Compensation of the phase modulation is only possible at the cost of decreased transmission. To exploit this trade-off, we have modified the criterion to the following form $E(t,r) = \theta_1 \cdot G + \theta_2 \cdot min(\langle T \rangle, T_m)$, where we have varied $T_m$ to optimize the phase with the constraint that the average transmission is not lower than $T_m$. Averaging is limited to the propagating part of the spatial spectrum. In this section, we use $\alpha = \beta = 1$ and $\theta_1 = \theta_2 = 1/2$.

In Figs. 10a, b, and c we present the optimization results for $T_m$ equal to *0.4*, *0.3*, and *0.2*, respectively, along with the transfer function obtained without any MDM at all (Fig. 10d). The phase of the transfer function becomes flat in the entire range only for $T_m \leq 0.2$. It is possible to compensate diffraction due to propagation at a larger distance than $\lambda$; however, the transmission is then rapidly suppressed.

The sub-wavelength imaging discussed in the previous section resembles more a projection than actual imaging. In contrast, compensation of Fresnel diffraction can be seen as imaging at a certain distance. Let us consider the propagation of a narrow Gaussian beam through the diffraction compensated MDM with $T_m = 0.2$. The width of the beam incident on the MDM becomes reconstructed at 200 nm from the other side of MDM. We illustrate this with FDTD simulation, shown in Figs. 10e and f.

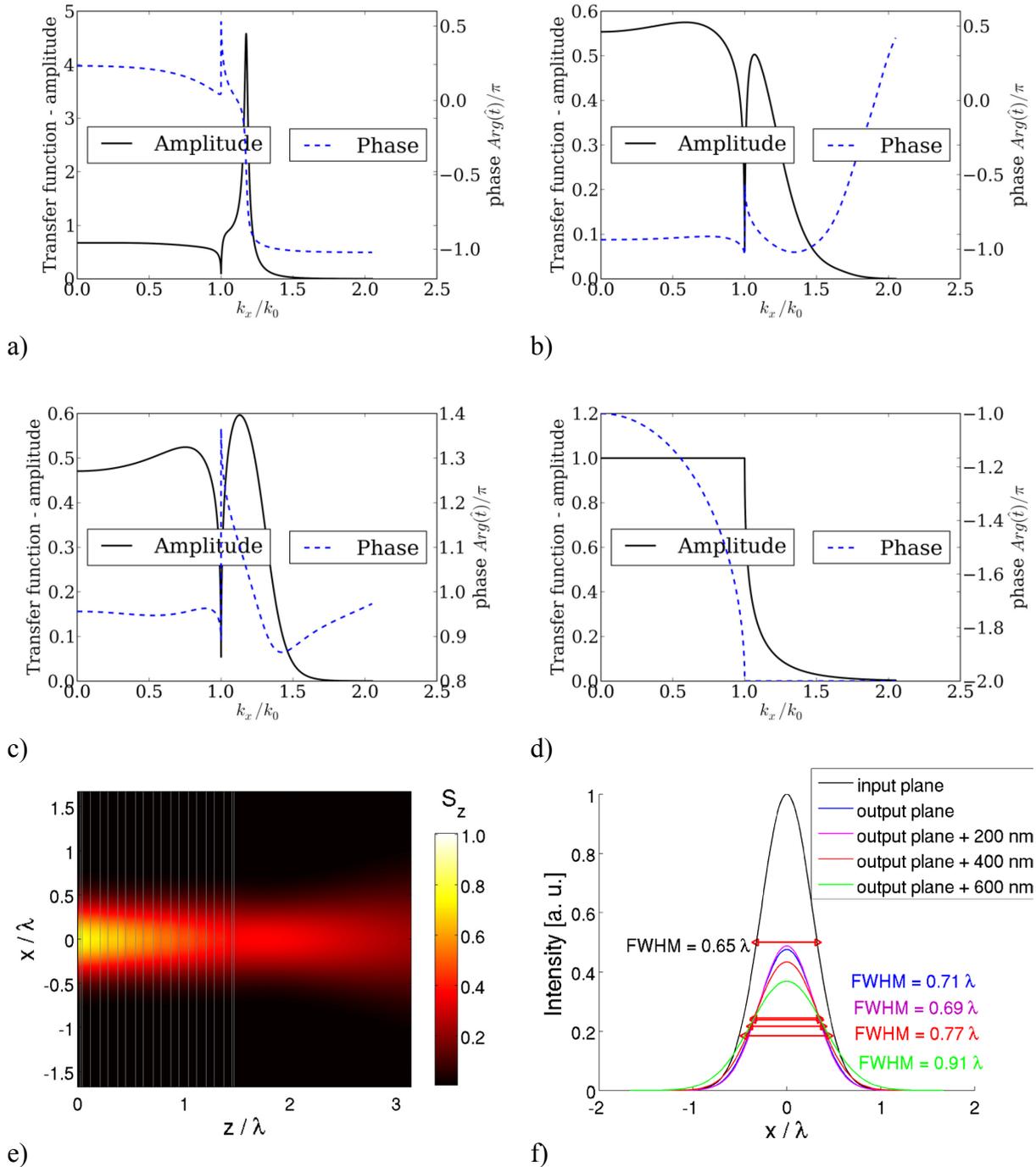

Figure 10: Transfer function of the MDM optimized for compensation of diffraction occurring at a distance of 200 nm. a) MDM obtained with the constraint $\langle T \rangle \geq 0.4$ defined with $d_0 = 5$ nm, $d_x = 42$ nm, $d_{Ag} = 48$ nm, $N = 2$, $\lambda = 413$ nm; b) MDM obtained with the constraint $\langle T \rangle \geq 0.3$ defined as $d_0 = 10$ nm, $d_x = 43$ nm, $d_{Ag} = 33$ nm, $N = 12$, $\lambda = 400$ nm; c) MDM obtained with the constraint $\langle T \rangle \geq 0.2$ defined as $d_0 = 10$ nm, $d_x = 45$ nm, $d_{Ag} = 35$ nm, $N = 8$, $\lambda = 400$ nm; d) diffraction at 200 nm without compensating MDM. e,f) FDTD simulation for the MDM from Fig.10d - the time-averaged Poynting vector component $S_z$ (e), and its cross-sections at various distances (f).

## 3.3 High-pass spatial filters for contrast change and for phase contrast

In this section we consider high-pass MDM spatial filters. One class of such filters may block the 0th spatial frequency, i.e. block transmission at normal incidence, and allow transmission for larger spatial frequencies. Such high-pass filters can be used to increase the contrast of an object, and to enhance its edges. An example of the transfer function of a high-pass filter is shown in Fig. 11a. Alternatively we could allow the 0th spatial frequency to go through; however, experiencing a different phase-shift than the higher frequencies. An example of such a filter is presented in Fig. 11b. Let us demonstrate its capability to enhance the edges of an object. This image processing operation is useful for microscopic imaging. For the purposes of example, the object is formed by an aperture in a metallic mask. Transmission through the aperture calculated with FDTD is shown in Fig. 11c. When the MDM is attached to the mask, the edges of the object become strongly enhanced, which is shown in Fig. 11d.

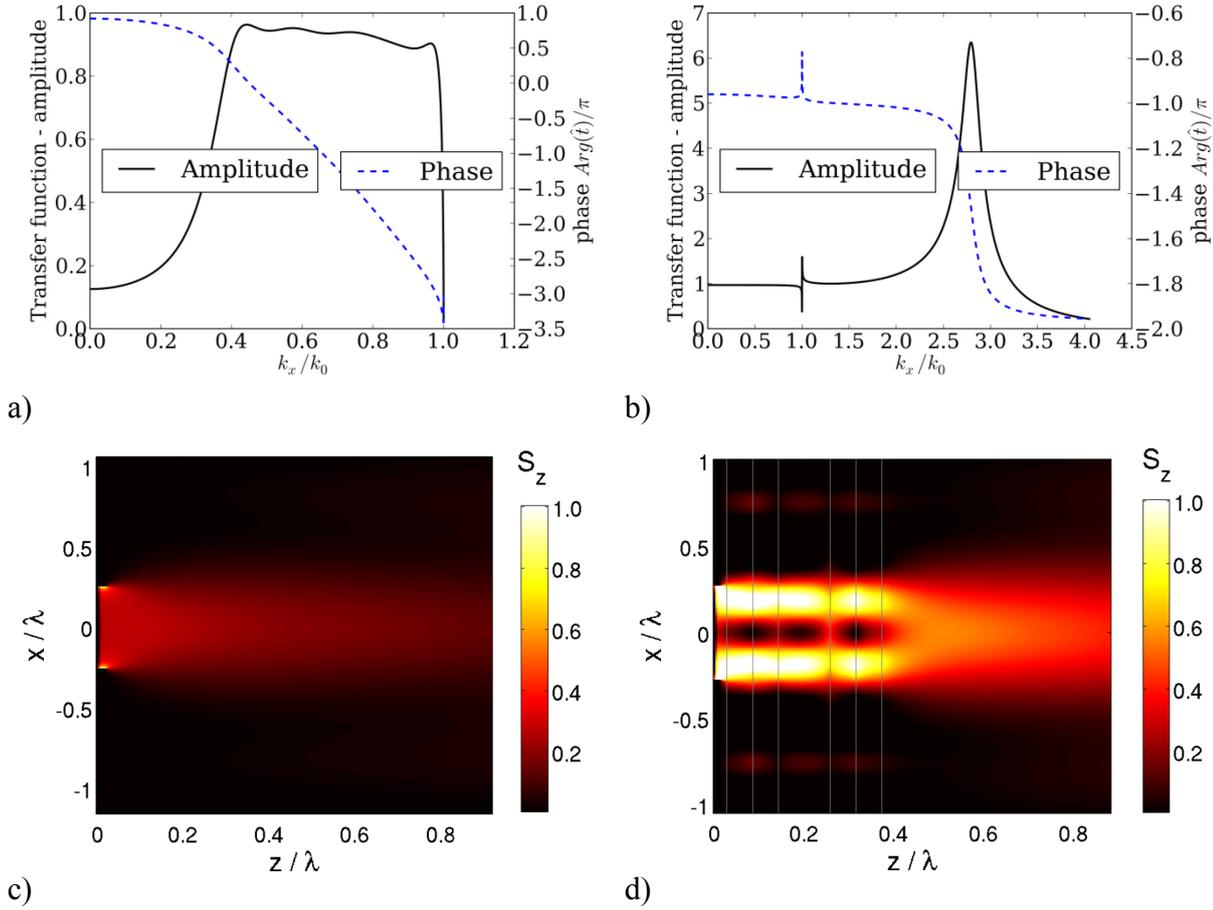

Figure 11: Transfer function of the MDM optimized for a) high-pass filtering $d_0 = 43$ nm, $d_x = 381$ nm, $d_{Ag} = 10$ nm, $N = 10, \lambda = 486$ nm; b) Sub-diffraction high-pass filters $d_0 = 24$ nm, $d_x = 48$ nm, $d_{Ag} = 24$ nm, $N = 1, \lambda = 455\ nm$; c,d) FDTD simulations with time-averaged Poynting vector component $S_z$ obtained for the transmission through an aperture in a metallic mask (c), and through the mask with an attached MDM same as in Fig. 11b.

## 4. Influence of surface roughness on linear filtering

Surface roughness has a strong influence on the expected functioning of plasmonic elements. The measured surface-plasmon–polariton (SPP) propagation lengths approach their theoretical values only with ultrasmooth pure metal films such as those obtained by combined template stripping with precisely patterned silicon substrates [32]. Recently, a silver superlens with smooth and low loss surfaces, capable of resolving objects with a resolution of $\lambda/12$, has been reported [33]. It was manufactured with nanoimprint technology and it contained an intermediate germanium wetting layer for the growth of flat silver films with surface roughness at subnanometer scales.

Resonant interactions in planar superlenses due to coupling between shadow mask features and surface roughness have been studied in [34]. Notably, surface roughness with RMS=5 nm of a thin silver film is sufficient to suppresses an SPP mode completely [35]. A detailed study of sub-wavelength imaging with MDM indicates that the tolerances to other parameters such as layer thickness or permittivity values are also critical [36]. Finally, in contrast to other reports, in [37] it has been found that a lens with periodic or random roughness can reduce field interference effects and provide improved focus on the transmission field.

In order to demonstrate the significance of surface roughness for imaging, we conducted an FDTD simulation of a silver-TiO$_2$ MDM with rough surfaces. The results shown in Fig. 12 include the time-averaged electromagnetic energy density. The MDM with layer thicknesses equal to $d_0 = 14,5$ nm, $d_x = 29$ nm, $d_{Ag} = 21$ nm with $N = 9$ periods operates at a wavelength of $\lambda = 427$ nm. The results clearly indicate that a subnanometer value of RMS is necessary to preserve the expected diffraction-free operation of the MDM. Surface roughness deteriorates transmission, creates hot spots and affects the resolution.

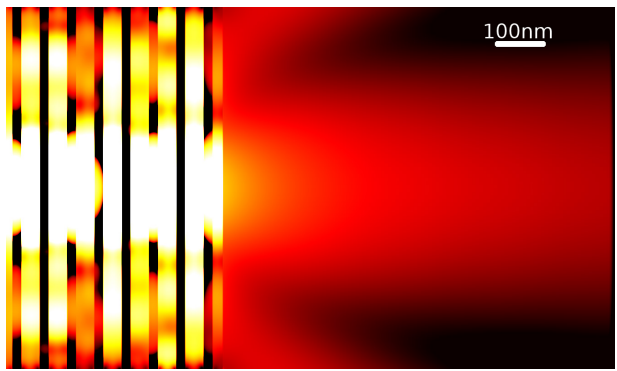

a)

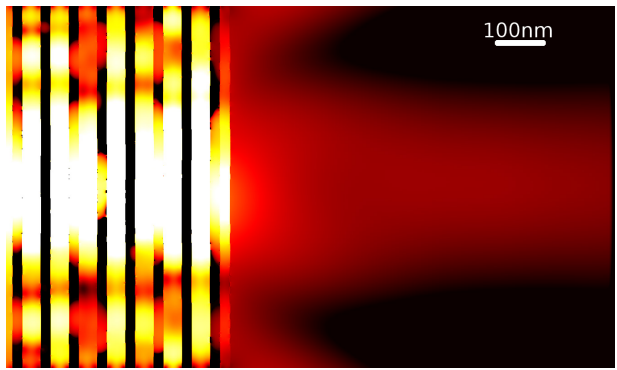

b)

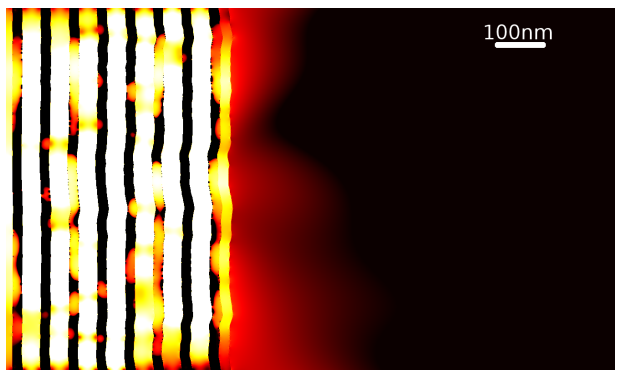

c)

Figure 12: Time-averaged energy density inside an MDM with rough layers, calculated with FDTD. a) RMS=0 nm, b) RMS=0.1 nm, c) RMS= 0.5 nm.

# 5. Conclusions

We reviewed the spatial filtering properties of MDMs. We also presented the modeling and optimization framework for engineering such metamaterials. A variety of point-spread functions may be reached, either for superresolution or for far-field image processing. The effective dispersion relation of the MDM had a profound effect on the shape of PSF.

We optimized the metamaterial with respect to the shape of the complex amplitude transfer function, the average transmission coefficient, and the average reflections. A measure of similarity obtained using Hölder's inequality was adapted to construct a criterion function.

Depending on the point spread function, the metamaterial could be applied for sub-diffraction spatial filtering or for far-field filtering operations on the wavefront.

# Acknowledgements

This work was supported by research project UMO-2011/01/B/ST3/02281 of the Polish National Science Center. PL-Grid infrastructure is acknowledged for providing access to computational resources.